\documentclass[prl,twocolumn,showpacs,amsmath,amssymb]{revtex4}
\usepackage{graphicx}

\begin{document}

\title{\boldmath
Anomalous Line-Shape of Cross Sections 
for $e^+e^- \rightarrow {\rm Hadrons}$
in the Center-of-Mass Energy Region between 3.650 and 3.872 GeV
}
\author{
\begin{small}
M.~Ablikim$^{1}$,              J.~Z.~Bai$^{1}$,               Y.~Ban$^{12}$,
X.~Cai$^{1}$,                  H.~F.~Chen$^{16}$,
H.~S.~Chen$^{1}$,              H.~X.~Chen$^{1}$,              J.~C.~Chen$^{1}$,
Jin~Chen$^{1}$,                Y.~B.~Chen$^{1}$
% ...
\end{small}
\\(BES Collaboration)\\
}
\vspace{0.2cm}
\affiliation{ 
\begin{minipage}{145mm}
$^{1}$ Institute of High Energy Physics, Beijing 100049, People's Republic of China\\
% ...
$^{12}$ Peking University, Beijing 100871, People's Republic of China\\
% ...
$^{16}$ University of Science and Technology of China, Hefei 230026, People's Republic of China\\
% ...
\end{minipage}
}
 
\begin{abstract}

We observe an obvious anomalous line-shape
of the $e^+e^- \rightarrow {\rm hadrons}$
total cross sections in the energy region between 3.700 and 3.872 GeV
from the data samples taken with the BES-II detector at the BEPC Collider.
Re-analysis of the data shows that it is inconsistent with the explanation 
for only one simple $\psi(3770)$ resonance with a statistical significance
of $7\sigma$.
The anomalous line-shape may be explained by two possible enhancements
of the inclusive hadron production near the center-of-mass energies of
3.764 GeV and 3.779 GeV, 
indicating 
that either there is likely
a new structure in addition to the $\psi(3770)$ resonance
around 3.773 GeV, or there are some physics effects
reflecting the $D\bar D$ production dynamics.
\end{abstract}

\maketitle

    In the energy range from 3.700 to 3.872 GeV, 
the well established $\psi(3770)$ resonance is 
believed to be the only observed structure.
This resonance has been identified to be a mixture of 
$D$-wave and $S$-wave of angular momentum
eigenstates of the $c\bar c$ system. In addition, the $\psi(3770)$ resonance 
is expected to decay into $D\bar D$ meson pairs 
with a branching fraction that is greater than $98\%$.
However, there is a long-standing puzzle in the existing measurements
of $\psi(3770)$ production and decays. 
Before recent
BES-II~\cite{bes2_psipp,prl97_121801_y2006,plb141_145_y2006,
bes_ekmax_prd76_122002_y2007,bes_ekmax_plb659_74_y2008} 
and CLEO-c~\cite{cleo_psipp} 
results published,
existing data indicated that about $38\%$ of $\psi(3770)$ does not decay to 
$D\bar D$ final states~\cite{rg_zdh_cjc}. Recently, the BES
Collaboration measured the branching fraction of $\psi(3770)$
decays to $D\bar D$ to be $B[\psi(3770)\rightarrow$
$D\bar D]=(85\pm 5)\%$     
~\cite{pdg07,prl97_121801_y2006,plb141_145_y2006}
and 
directly measured the non-$D\bar D$ branching fraction
of $\psi(3770)$ decay to be
$B[\psi(3770)\rightarrow$non-$D\bar D]=
   (13.4 \pm 5.0 \pm 3.6)\%$~\cite{bes_ekmax_prd76_122002_y2007} and
$B[\psi(3770)\rightarrow$non-$D\bar D]=
(15.1 \pm 5.6 \pm 1.8)\%$~\cite{bes_ekmax_plb659_74_y2008}
under assumption that there is only one simple $\psi(3770)$ resonance in the
energy region between 3.700 and 3.872 GeV.
In the last two years, the BES and CLEO Collaborations have searched for 
exclusive non-$D\bar D$ decays of $\psi(3770)$.
However, the summed non-$D\bar D$ branching fraction measured by both the BES
and CLEO Collaborations remains to be less than
$2\%$~\cite{bes2_psipp,cleo_psipp}.
To understand why the measured inclusive non-$D\bar D$ 
branching fraction is substantially larger than $2\%$,
in addition to continuing searching for more possible non-$D\bar D$ 
decay modes of $\psi(3770)$, it is worth going back to carefully examine 
the previous measurements of the $\psi(3770)$ parameters. 

     An examination of analyses previously reported by the BES Collaboration in
Refs.~\cite{prl97_121801_y2006,plb652_238_y2007} shows that
the fits to the observed hadronic cross sections or $R$ values
are rather poor for the fine-grained energy scan 
cross section measurements
[see Fig. 4(a)
in Ref.~\cite{prl97_121801_y2006}
and Fig. 1 in Ref.~\cite{plb652_238_y2007}] 
even though the branching fraction for $\psi(3770)\rightarrow
$non-$D\bar D$ was left as a free parameter in the fits. 
In this letter, we present a reanalysis of the observed 
inclusive hadronic cross sections to better understand 
the hadronic annihilation structure in the energy region
between 3.700 and 3.872 GeV.

\begin{figure}[hbt]
\includegraphics*[width=8.3cm,height=7.0cm]  
{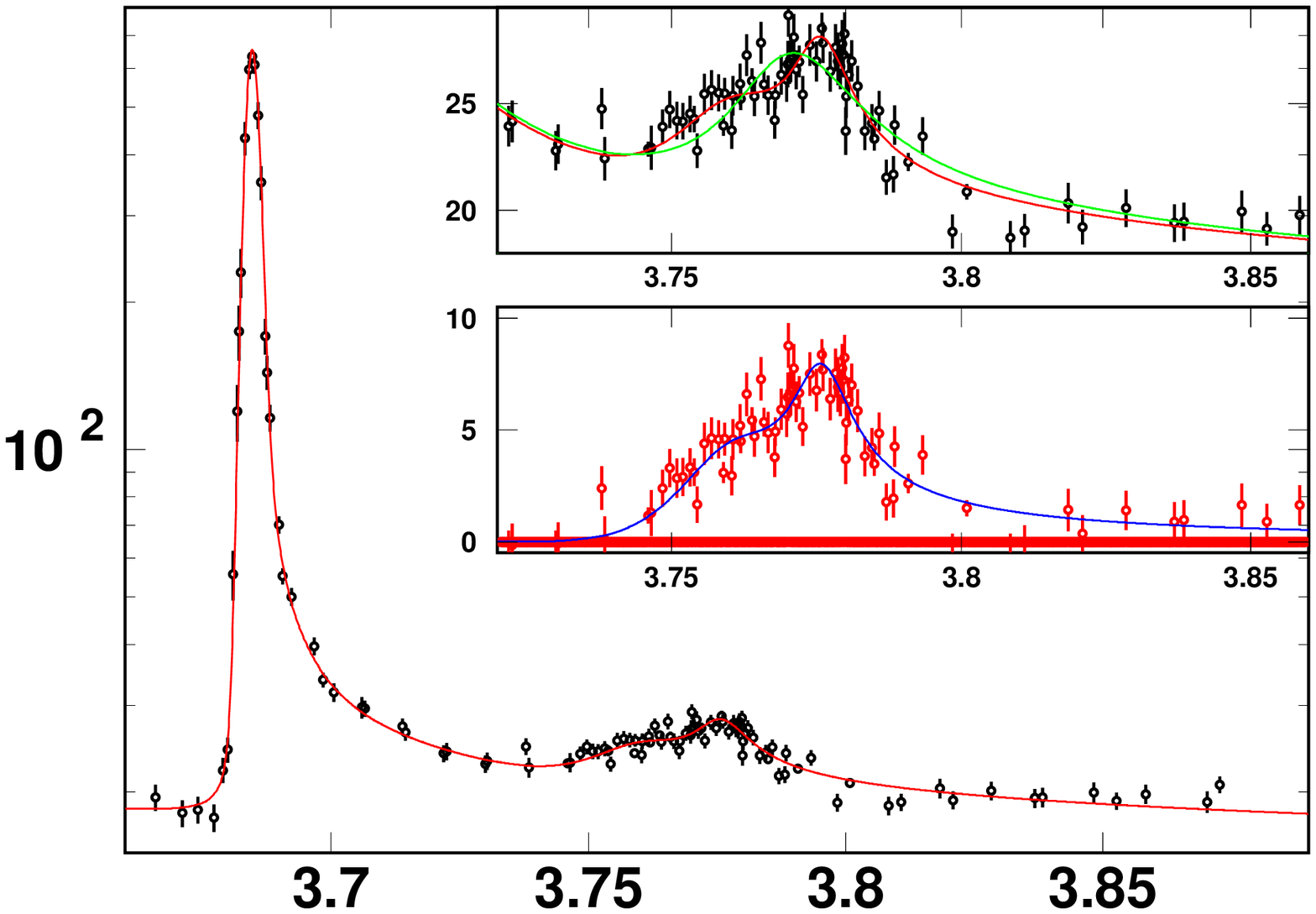}
\put(-155,-0){\bf\large $E_{\rm cm}$ [GeV]}
\put(-250,75){\rotatebox{90}{\bf\large {$\sigma^{\rm obs}_{\rm
had}$~~~[nb]}}}
\put(-30,175){\large (a)}
\put(-30,115){\large (b)} 
\caption{
The measured inclusive hadronic cross sections versus the
c.m. energy for the two data sets taken in March and December 2003;
the fit is done with two incoherent amplitudes (solution 1), see text for
detail.
}
\label{fg:xsct_49pnts_67pnts_noitf}
\end{figure}

     The measurements of the observed inclusive hadronic cross sections 
are discussed in detail in the
Refs.~\cite{prl97_121801_y2006, plb652_238_y2007,
plb141_145_y2006,prl97_262001_y2006}.
The observed inclusive hadronic cross sections obtained from
the cross section scan data taken in March 2003 and in December 2003
are illustrated 
in Fig.~\ref{fg:xsct_49pnts_67pnts_noitf} 
~\footnote{The observed cross sections from the two data sets are corrected
with the measured $R_{\rm uds}$ values before combining them together.} 
by dot with error bars,
where the error bars are the combined statistical and point-to-point systematic
uncertainties. 
The systematic uncertainty
includes the statistical uncertainty of the luminosity, the
uncertainties of the Monte Carlo efficiencies for detections of the Bhabha
scattering events and the hadronic events, 
as well as the uncertainty of the observed cross sections due to the
reproducibility ($\pm 0.1$ MeV)
of setting the BEPC machine energy.
The c.m. (center-of-mass) energy of the BEPC machine
is calibrated 
with the world average masses of $\psi(3686)$ and $J/\psi$.
The measured masses of $\psi(3686)$ and $J/\psi$ at BEPC
are obtained by analyzing 6 data sets of $\psi(3686)$ scan
and 2 data sets of $J/\psi$ scan
performed during the time periods of collecting the finer
cross section scan data.
The uncertainty in the calibrated energy for
the combined two finer cross section scan data sets together 
is about $\pm 0.5$ MeV.

A close examination of the energy region 
(from 3.74 to 3.80 GeV) around 3.777 GeV 
shows that the slopes of the observed cross sections 
on the two sides of the peak 
are quite different; with the slope of the high energy side
of the peak substantially 
larger than that of the low energy side.
It conflicts with the expectations for
only one resonance in this energy region,
since the effects of the initial state radiation (ISR) 
and the $D\bar D$ production threshold 
as well as the energy dependence of the $D\bar D$ scattering amplitudes
due to the Blatt-Weisskopt barrier ~\cite{Blatt-Weisskopt_1962}
would all make 
the slope at the high energy side of the peak less steep relative to the
slope on the low-energy side.
This anomalous shape seen in the precision measurement 
indicates that one simple resonance hypothesis is quite questionable 
to fit the current data. 
Instead of the conventional definition 
of the $\psi(3770)$ decay width $\Gamma(E_{\rm cm})$, 
if the dynamics of
$D\bar D$ scattering or some reasonable model describing the $D\bar D$
scattering can give some special form of $\Gamma(E_{\rm cm})$ and mass shift
for which the scattering amplitude gets zero or node at the rather low $D$  
meson momentum ($P_D\sim 0.4$ GeV) to adapt the unusal decline around 3.8 GeV 
in the cross section line shape, the anomalous line-shape of the cross
sections for $e^+e^-\rightarrow {\rm hadrons}$ might be understood. 

However, as shown in this work, it can not be excluded 
neither
that there may be some new structure 
in addition to the $\psi(3770)$ resonance in
the energy region between 3.700 and 3.872 GeV,
which and its interference with
the $\psi(3770)$ amplitude distort the line-shape of the observed
cross section from that expected if there was only one resonance in the
region.

To investigate whether there are some new structures in addition to the $\psi(3770)$
resonance in the energy region between 3.700 and 3.872 GeV, we fit the
observed cross sections with one or two amplitudes in the energy region.
The expected cross section $\sigma^{\rm expect}_{\rm had}(E_{\rm cm})$
consisting of four components can be given as
\begin{eqnarray}
\sigma^{\rm expect}_{\rm had}(E_{\rm cm}) =
    \sigma^{\rm expect}_{Rs(3770)}(E_{\rm cm})
  + \sigma^{\rm expect}_{J/\psi}(E_{\rm cm}) \nonumber \\  
  \hspace{6mm} + \sigma^{\rm expect}_{\psi(3686)}(E_{\rm cm})
  + \sigma^{\rm CTM}_{\rm had}(E_{\rm cm}),
\end{eqnarray}
in which
$\sigma^{\rm expect}_{Rs(3770)}(E_{\rm cm})$, 
$\sigma^{\rm expect}_{J/\psi}(E_{\rm cm})$, 
$\sigma^{\rm expect}_{\psi(3686)}(E_{\rm cm})$,
and 
$\sigma^{\rm CTM}_{\rm had}(E_{\rm cm})$
are, respectively,
the expected cross sections for
$Rs(3770) \rightarrow {\rm hadrons}$,
$J/\psi \rightarrow {\rm hadrons}$, $\psi(3686) \rightarrow {\rm hadrons}$,
and continuum light hadron
production at the c.m. energy $E_{\rm cm}$, 
and $Rs(3770)$ denotes
the full structure around 3.773 GeV.
The expected cross sections are obtained from the Born order
cross sections for these processes and the ISR
corrections~\cite{kuraev, berends}. 

For the $Rs(3770)$ resonance(s), we use one or two pure
P-wave Breit-Wigner amplitude(s)
with energy-dependent total
widths~\cite{prl97_121801_y2006,plb141_145_y2006,plb652_238_y2007}
to fit the observed hadronic cross sections.
The two amplitudes are expected as
\begin{equation}
A_j(E_{\rm cm}) = \frac{ \sqrt{ 12\pi \Gamma^{ee}_j \Gamma^{\rm had}_j }}
           {(E_{\rm cm}^2-M_j^2) + i \Gamma^{\rm tot}_j(E_{\rm cm}) M_j }~~(j=1,2),
\end{equation}
where
$M_j$, $\Gamma^{ee}_j$,
$\Gamma^{\rm had}_j$, 
and $\Gamma^{\rm tot}_j(s)$
are the masses, leptonic widths, hadronic
widths,
and the total widths of the two resonances, respectively.
$\Gamma^{\rm tot}_j(E_{\rm cm})$ is
chosen to be energy
dependent~\cite{prl97_121801_y2006,plb141_145_y2006,plb652_238_y2007}.
For two amplitude hypothesis, concerning the possible interference 
between the two amplitudes, 
we use 
two extreme schemes
to see if we can get better description for
the anomalous line shape.
In the first scheme,
we ignore the possible interference; and in the second,
we assume the complete interference between the two 
amplitudes. These two schemes
give the Solution 1 and Solution 2, respectively.
The Born order cross section 
for $Rs(3770)$ production in Solution 1 and Solution 2 can, respectively,
be written as
\begin{equation}  
\sigma_{Rs(3770)}(E_{\rm cm}) = |A_1(E_{\rm cm})|^2+|A_2(E_{\rm cm})|^2
\end{equation}  
and
\begin{equation}  
\sigma_{Rs(3770)}(E_{\rm cm}) = |A_1(E_{\rm cm})+ e^{i\phi}A_2(E_{\rm
cm})|^2,
\end{equation}  
where the $\phi$ is the relative phase difference
between the two amplitudes.

The non-resonant background shape is taken as
\begin{equation}
\sigma^{\rm CTM}_{\rm had}(E_{\rm cm}) = \sigma^{\rm CTM}_{\rm LtHd}(E_{\rm cm}) 
                      + \sigma^{\rm CTM}_{D\bar D}(E_{\rm cm})
\end{equation}
\noindent
with
\begin{equation}
\sigma^{\rm CTM}_{D\bar D}(E_{\rm cm})=
 f\left[ (\frac{p_{D^0}}{E_{D^0}})^3\theta_{00}
                  + (\frac{p_{D^+}}{E_{D^+}})^3\theta_{+-} \right]
 \sigma^B_{\mu^+\mu^-}(E_{\rm cm}),
\end{equation}
\noindent
where $\sigma^{\rm CTM}_{\rm LtHd}(E_{\rm cm})$ is the observed cross section for
light hadronic event production given in
Refs.~\cite{prl97_121801_y2006, plb652_238_y2007},
$\sigma^B_{\mu^+\mu^-}(s)$ is the Born cross section
for $e^+e^-\rightarrow \mu^+\mu^-$, $p_{D^0}$ and $p_{D^+}$
($E_{D^0}$ and $E_{D^+}$) are the momenta (energies) of $D^0$ and $D^+$ mesons
produced at the nominal energy $\sqrt{s}$, $\theta_{00}$ and $\theta_{+-}$
are the step functions to account for the thresholds of the $D^0\bar D^0$ and
$D^+D^-$ meson pair production, respectively;
$f$ is a parameter to be fitted.
The effect of energy spread on the observed cross sections is also
considered in the analysis.    

\begin{figure}[hbt]
\includegraphics*[width=8.3cm,height=7.0cm]   
{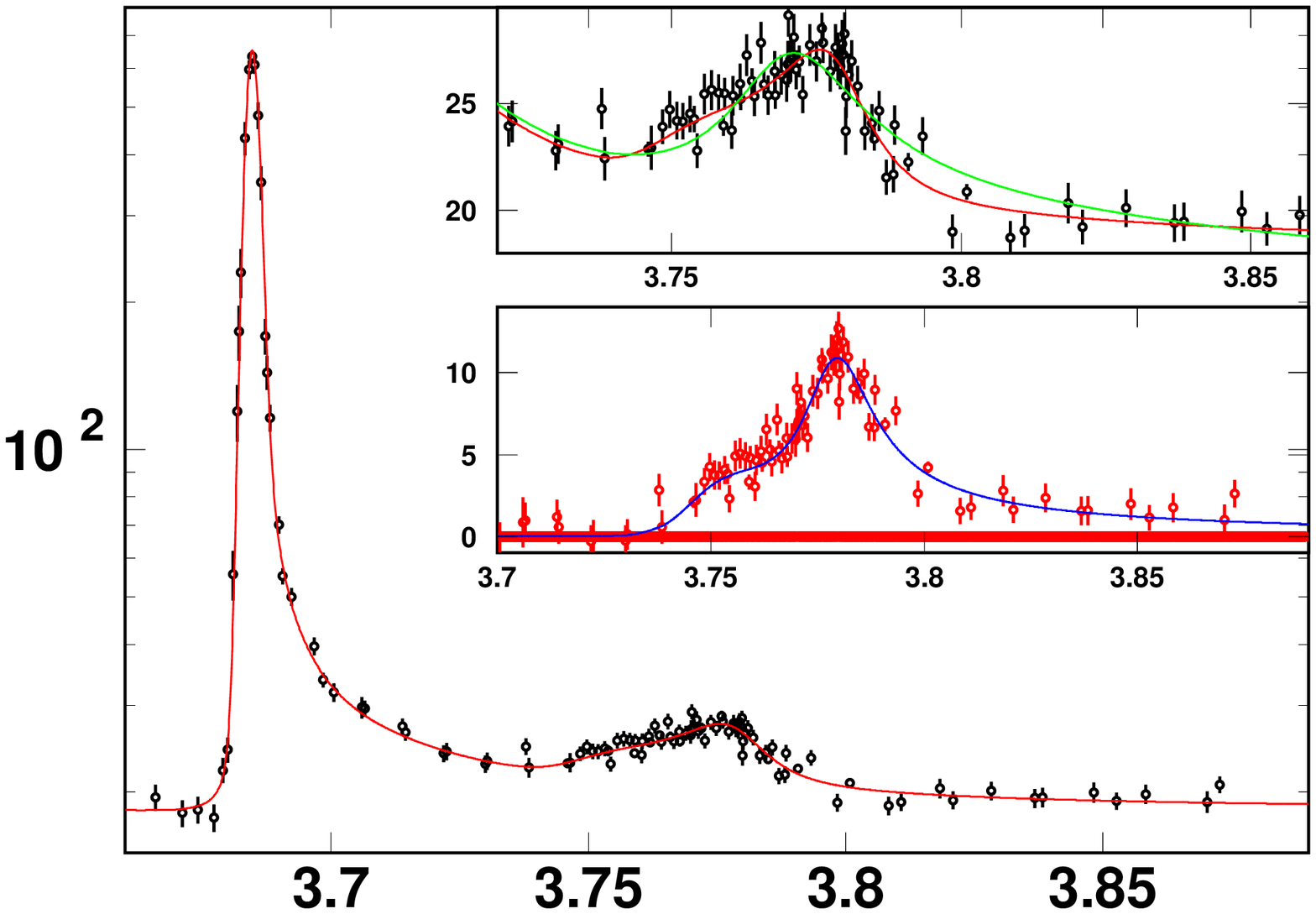}
\put(-155,-0){\bf\large $E_{\rm cm}$ [GeV]} 
\put(-250,75){\rotatebox{90}{\bf\large {$\sigma^{\rm obs}_{\rm
had}$~~~[nb]}}}
\put(-30,175){\large (a)}
\put(-30,115){\large (b)}
\caption{
The observed inclusive hadronic cross sections versus the nominal
c.m. energies for the combined data sets taken in March and December 2003;
the fit was done with two coherent amplitudes for $Rs(3770)$ (Solution 2).  
}
\label{fg:xsct_49pnts_67points}
\end{figure}

In the following, 
ignoring the tiny difference of the detection efficiencies determined from
the different schemes 
as described above,
we fit the observed cross sections presented in
Fig.~\ref{fg:xsct_49pnts_67pnts_noitf} and
in Fig.~\ref{fg:xsct_49pnts_67points}, respectively, with the 
expected cross sections given in Eq. (1) in two schemes. 
In the first case, it is defined in Eq. (3)
and the fits give the results of the Solution 1. In the second case,
it is defined in the Eq.(4) and the fit gives
the results of the Solution 2. 
As a comparison we also fit the cross sections with the conventional 
one Briet-Wigner form of $\psi(3770)$ resonance as the definition of
the $R_s(3770)$ for the one resonance hypothesis.
In the fits, 
we fix $r=1.5$ fm ($r$ is the interaction radius of the 
$c\bar c$ system)~\cite{prl97_121801_y2006,plb141_145_y2006,plb652_238_y2007}
and fix the $J/\psi$ parameters at the values given in PDG07~\cite{pdg07};   
the $\psi(3686)$ and $Rs(3770)$ resonance parameters are left free, 
$R_{\rm uds}$ and $f$~\cite{prl97_121801_y2006, plb652_238_y2007}   
are also left free. 
\begin{table*}[t]
\caption{The fitted results for
the data taken in March 2003 and December 2003.}
\label{tbl:two_amplitudes_er015_49pnts_67pnts}  
\begin{tabular}{lcccr} \hline \hline
Quantity &   two amplitudes       &  two amplitudes &  one amplitude &
$\psi(3770)$ and $G(3900)$ amplitudes\\
  & (without interference) &  (interference) &  & (interference) \\
  &  Solution 1  &  Solution 2   & & Solution 3 \\  \hline
$\chi^2/ndof$  & $125/103=1.21$  & $112/102=1.10$ &  $182/106=1.72$ &
$170/104=1.63$  \\
$M_{\psi(3686)}$ [MeV] & $3685.5\pm 0.0 \pm 0.5$
                       & $3685.5\pm 0.0 \pm 0.5$ &  $3685.5\pm 0.0\pm 0.5$
                       & $3685.5\pm 0.0 \pm 0.5$ \\
$\Gamma^{\rm tot}_{\psi(3686)}$ [keV]  & $312\pm 34\pm 1$
                        & $311\pm 38\pm 1$ &   $304 \pm 36 \pm 1 $
                        & $293\pm 36\pm 1$ \\
$\Gamma^{ee}_{\psi(3686)}$ [keV] & $2.24\pm 0.04\pm 0.11$
                                 & $2.23\pm 0.04\pm 0.11$
                                 & $2.24\pm 0.04\pm 0.11$
                                 & $2.23\pm 0.04\pm 0.11$ \\
$M_1$ [MeV]                     & $3765.0\pm 2.4 \pm 0.5$   
                                & $3762.6\pm 11.8 \pm 0.5$   
                                & $3773.3\pm 0.5 \pm 0.5$   
                                & $3774.4\pm 0.5 \pm 0.5$   \\
$\Gamma^{\rm tot}_{1}$ [MeV] & $28.5\pm 4.6\pm 0.1$  
                             & $49.9\pm 32.1\pm 0.1$ 
                             & $28.2 \pm 2.1\pm 0.1 $
                             & $28.6\pm 2.3\pm 0.1$  \\
$\Gamma^{ee}_{1}$ [eV] & $155\pm 34\pm 8$
                       & $186\pm 201\pm 8$
                       & $260 \pm 21 \pm 8 $
                       & $264 \pm 23\pm 8$ \\
$M_2$ [MeV]            & $3777.0\pm 0.6 \pm 0.5$ &  $3781.0\pm 1.3 \pm 0.5$
                       &  -- 
                       & $3943.0$ (fixed) \\
$\Gamma^{\rm tot}_{2}$ [MeV]   
                       & $12.3\pm 2.4 \pm 0.1$ & $19.3\pm 3.1 \pm 0.1$
                             & --
                             & -- \\
~~~~or $\sigma_G$ [MeV]   & -- &-- & -- & $54$ (fixed) \\

$\Gamma^{ee}_{2}$ [eV]
                       & $93 \pm 26\pm 9$   & $243 \pm 160 \pm 9$
                       & --
                       & --- \\
~~~~or C               & ---   & -- & -- & $0.243$ (fixed) \\

$\phi$ [$^o$]          &     --
                       &    $(158 \pm 334 \pm 5)$
                       &   --
                       & $(150 \pm 23 \pm 5)$   \\
   $f$                & $0.4\pm 5.6\pm 0.6$
                      & $5.2\pm 2.5\pm 0.6$
                      & $0.0\pm 0.5\pm 0.6$
                      & $0.0\pm 1.2\pm 0.6$ \\
\hline \hline
\end{tabular}
\end{table*} 

As shown in Fig.~\ref{fg:xsct_49pnts_67pnts_noitf}
and in Fig.~\ref{fg:xsct_49pnts_67points},
the circles with error bars show the observed cross sections. The red lines 
in both of the figures
and in the sub-figures (a) inserted in
Fig.~\ref{fg:xsct_49pnts_67pnts_noitf}
and Fig.~\ref{fg:xsct_49pnts_67points}
represent the fitted values of the cross sections
of Solution 1 and Solution 2.
The green lines in the sub-figures (a)
show the fit to the observed cross sections
for the one amplitude hypothesis.
The circles with error bars in red 
as shown in the sub-figuares (b) inserted in 
Fig.~\ref{fg:xsct_49pnts_67pnts_noitf}
and Fig.~\ref{fg:xsct_49pnts_67points} show the
measured {\it net cross sections}, which are obtained
by subtracting the contributions from $J/\psi$ and $\psi(3686)$ decays
to hadrons, the continuum hadron production and the interference term 
of the two amplitudes in $Rs(3770)$ definition for the Solution 2;
the blue lines show the fit to the {\it net cross sections} from
the two resonances for both of the Solution 1 and Solution 2, respectively.

The 2$nd$, the 3$rd$ and the 4$th$ columns
of Table~\ref{tbl:two_amplitudes_er015_49pnts_67pnts}
summarize, respectively, the results of the fits for the Solution 1 and
the Solution 2 of the two amplitude hypothesis, and
for the one amplitude hypothesis,
where the first errors are from the fit and the second systematic.
For the measured masses, the second errors mainly arise from the uncertainty
of the BEPC machine energy calibration for the combined two data sets together.
For the one resonance hypothesis, 
the fit yields $\psi(3770)$ and $\psi(3686)$
parameters as listed in the 4$th$ column of
Table~\ref{tbl:two_amplitudes_er015_49pnts_67pnts}.
These measured values of the 
resonance parameters
are consistent within error with the world averages
~\cite{pdg06}~\footnote{We compare our results with
PDG06 world average, since PDG06 did not include BES results on the
measurements}
and with the earlier BES measurements~\cite{prl97_121801_y2006}~\cite{
plb141_145_y2006} obtained by analyzing the two data samples separately.
The fit gives the mass difference between the $\psi(3770)$ and $\psi(3686)$
resonances to be $\Delta_M=87.8\pm 0.5$ MeV. However, the large $\chi^2/ndof$
in the 4$th$ column of
Table~\ref{tbl:two_amplitudes_er015_49pnts_67pnts} gives the fit probability
of less than $7\times 10^{-6}$, 
meaning that the one resonance hypothesis is strongly
incomparable with the present precision measurement data.
On the contrary,
the $\chi^2$ change for the two hypotheses in Solution 1 
is $(182-125)=57$ with a reduction of 3 degrees of freedom.
This indicates that the signal significance for the new structure 
is $7.0\sigma$.
The $\chi^2$ change for the two hypotheses in the Solution 2 is 
$70$ with a reduction of 4 degrees of freedom.
This indicates that the statistical significance
of the new structure is $7.6\sigma$.
Comparing the fits for the Solution 1 and Solution 2, 
we find that the $\chi^2$ change of $13$ with a reduction of
1 degree of freedom.
The significance of the interference 
between the two Breit-Wigner amplitudes is $3.6\sigma$,
which indicates that the two
amplitudes likely interfere somehow with each other.
The actual situation of the interference
would be somewhere between the two cases. It depends on
what are the exact final states
of the possible new structure decays.

\begin{figure}[hbt]
\includegraphics*[width=8.3cm,height=7.0cm]
{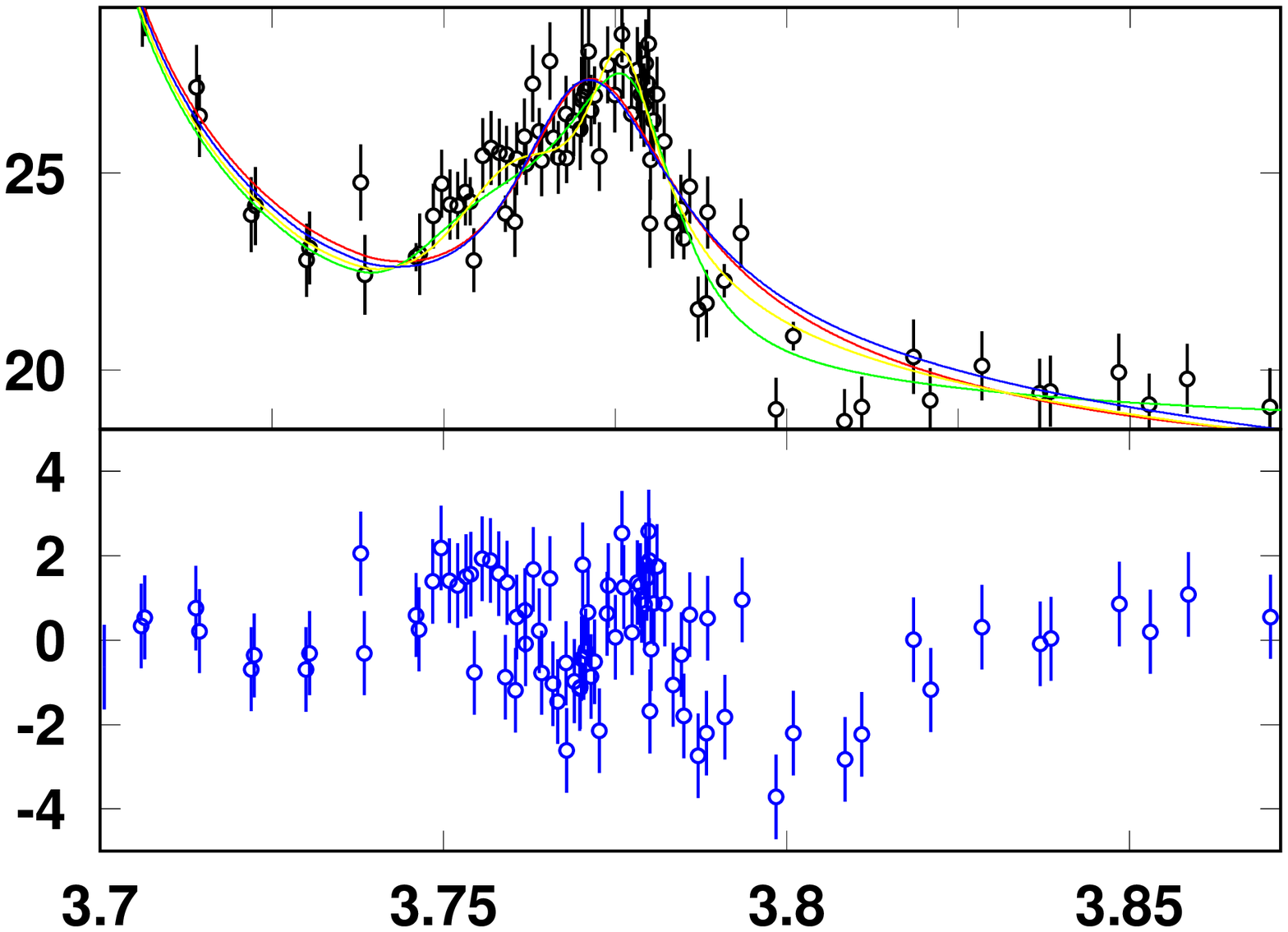}
\put(-155,-0){\bf\large $E_{\rm cm}$ [GeV]}
\put(-245,125){\rotatebox{90}{\bf\large {$\sigma^{\rm obs}_{\rm
had}$~[nb]}}}
\put(-240,55){\rotatebox{90}{\large {Ratio}}}
\put(-50,155){\large (a)}
\put(-50,85){\large (b)} 
\caption{
(a) the observed inclusive hadronic cross section versus the nominal
c.m. energy; (b) ratio of residual to error of observed cross section; (see text).
}
\label{fg:xsct_49pnts_67points_psi3780_g3940}
\end{figure}

However, it is noted that the fitted value
$f=5.2\pm 2.5\pm 0.6$ in the Solution 2         
would lead to a huge $D\bar D$ production cross section 
at higher energy region and there exists an evident dip of the
inclusive hadronic cross section around $E_{cm}$=3.80 GeV. 
These indicate that, instead of only the continuum $D\bar D$ production,
there might be a broad structure whose peak
is at higher energy than 3.83
GeV and it interferes with $Rs(3770)$. Recently, BABAR~\cite{3900_BABAR} and BELLE
collaborations~\cite{3900_BELLE} observed $G(3900)$.
To consider the effect of the $G(3900)$ on the observed
cross sections,
instead of the first two 
solutions for the two structure hypotheses one may adopt the third
approach by including the new component of $D\bar D$ production amplitude 
of $G(3900)$.
The fitting procedure is analogous to 
Solution 2. 
However, the amplitude $A_2(E_{cm})$ in Eq. (4) is replaced by
a square root product of a parameter C and a Gaussian function $G$.
The mass and the standard deviation of $G$ are fixed at the measured values 
of $G(3900)$~\cite{3900_BABAR} and $C$ is fixed 
at 0.243 corresponding to the $D\bar D$ cross section
as the one measured by BABAR  at 3.943 GeV.
The red line in Fig.~\ref{fg:xsct_49pnts_67points_psi3780_g3940}(a)
represents the fitted values of the cross sections,
which is obtained from the
fit under assumption that 
the $\psi(3770)$ and $G(3900)$ amplitudes interfere with each other;
the fitted value from the hypotheses for only $\psi(3770)$ amplitude (blue
line), 
from Solution 1 (yellow line) and from Solution 2 (green line) 
are also illustrated in
Fig.~\ref{fg:xsct_49pnts_67points_psi3780_g3940}(a).
The 5$th$ column of Table~\ref{tbl:two_amplitudes_er015_49pnts_67pnts}
summarizes the results (Solution 3) of the fit 
including $G(3900)$.
The fit gives a rather poor
fit probability of less than $5\times 10^{-5}$, 
which does not significantly improve
the fit from the one resonance hypothesis.
If we consider three coherent amplitudes in the fit by replacing
$|A_1(E_{cm})+e^{i\phi}A_2(E_{cm})|^2$ with
$|A_1(E_{cm})+e^{i\phi_1}A_2(E_{cm})+e^{i\phi_2}G(E_{cm})|^2$ in Eq. (4),
where $G$ is the $G(3900)$ structure, 
we obtain almost the same results as these shown in Solution 2 in Table I
instead of $f=5.2 \pm 2.5 \pm 0.6$. This fit gives
$f=2.7 \pm 6.4 \pm 0.6$, which is comparable 
with the inclusive hadronic cross section measurements at the
higher energy region.
Fig.~\ref{fg:xsct_49pnts_67points_psi3780_g3940}(b) shows
the ratio of the residual between the observed cross section and the
fitted value for the one $\psi(3770)$ amplitude hypothesis to the error of
the observed cross section.
The variation of the ratio with $E_{\rm cm}$ indicates that there is more
likely some new structure additional to $\psi(3770)$ resonance.

In summary, 
by re-analyzing the line-shape of the cross sections for
$e^+e^-\rightarrow {\rm hadrons}$, 
we find that it does not describe the cross section shape well with
the hypothese that 
only one simple $\psi(3770)$ resonance exists  
in the energy region from 3.700 to 3.872 GeV. 
If there are no other dynamics effects which distort the pure D-wave Breit-Weigner
shape of the cross sections, 
the analysis shows that the fit is inconsistent with the explanation for
only one simple $\psi(3770)$ resonance there at $7\sigma$
statistical significance,
indicating that there might be evidence for a new structure additional
to the single $\psi(3770)$ resonance. 
However, if there are some dynamics effects distorting the pure D-wave
Breit-Weigner shape of the cross sections, such as the 
rescattering of $D\bar D$ leading to
the significant energy dependence of the wave function in the
$D\bar D$ decays
of the $\psi(3770)$ resonance,
one has to consider those
effects in the measurements of the resonance parameters of
$\psi(3770)$,
since these effects would
definitely shift the measured values of the resonance parameters.
Anyway, the large non-$D\bar D$ branching fraction of $\psi(3770)$ decays measured
previously~\cite{prl97_121801_y2006,plb141_145_y2006} 
may partially be due to the assumption that there is only
one simple resonance in the energy region between 3.700 and 3.872 GeV in
the previous measurements of the $\psi(3770)$ parameters.

   The BES collaboration thanks the staff of BEPC for their hard efforts.
This work is supported in part by the National Natural Science Foundation
of China under contracts
Nos. 19991480,10225524,10225525, the Chinese Academy
of Sciences under contract No. KJ 95T-03, the 100 Talents Program of CAS
under Contract Nos. U-11, U-24, U-25, and the Knowledge Innovation Project
of CAS under Contract Nos. U-602, U-34(IHEP); by the
National Natural Science
Foundation of China under Contract No.10175060(USTC),and
No.10225522(Tsinghua University).

\end{document}